\documentclass[12pt]{iopart}
\usepackage{iopams}
\begin{document}
\title{ A relativistically covariant version of Bohm's quantum field theory for the scalar field  }
\author{George Horton and Chris Dewdney}
\address{Division of Physics, University of Portsmouth. Portsmouth PO1 2DT. England}

\begin{abstract}

We give a relativistically covariant, wave-functional formulation
of Bohm's quantum field theory for the scalar field based on a
general foliation of space-time by space-like hypersurfaces. The
wave functional, which guides the evolution of the field, is
space-time-foliation independent but the field itself is not.
Hence, in order to have a theory in which the field may be
considered a beable, some extra rule must be given to determine
the foliation. We suggest one such rule based on the eigen vectors
of the energy-momentum tensor of the field itself.
\end{abstract}
\section{Introduction}

In this paper we begin by deriving a suitable expression for a
scalar Hamiltonian of a real classical scalar field on a
space-like hypersurface. This is an essential difference from
Bohm's starting point which takes the time component of the total
four-momentum as the Hamiltonian and implicitly assumes the space
components are zero. This implicitly introduces a preferred frame
of reference and hence a non-covariant theory. It is therefore not
surprising that Bohm finds the vacuum field to be stationary in
one particular frame only. In addition, an arbitrary foliation of
flat space-time by the equal-time hyperplanes of the chosen frame
was tacitly assumed. Such a procedure obviously conflicts with the
requirements of a relativistically invariant theory. Formulating
the theory in a fully covariant manner makes apparent the
dependence of the field evolution on the foliation of space-time
and emphasizes that this has nothing to do with the arbitrary
frame of reference used to describe the system. We proceed to
quantise the field on a general hypersurface and to develop the
form of the wave functionals of the quantised field. We show how
to apply the Hamiltonian density to the wave functional and give
two explicit examples. We then demonstrate how one may integrate
the equations of motion to calculate the evolution of the Bohm
field over an arbitrary family of space-like hypersurfaces. Noting
that the foliation-dependence of the field is somewhat at odds
with the desire to produce a theory of beables, it is clear that
some extra rule is required to determine the foliation
\footnote{Duerr et al \cite{Duerr1999} have previously discussed
the need for some additional structure to determine the required
foliation of space-time in the context of their Bohm-Dirac
particle trajectory model.}. We propose that, given some
``initial" space-like hypersurface, and the field thereon, the
foliation may be determined in a natural way using the flows of
energy-momentum determined by the field itself.
\section{Covariant Formulation of Quantum Field Theory}
As shown by Schwinger \cite{Schwinger1948} the evolution of a
state vector $\Psi[\sigma]$ on a space-like hypersurface $\sigma$
is given by an equation of the form
\begin{equation}\label{eq:Schwinger}
i\frac{\delta \Psi[\sigma]}{\delta\sigma(x)}=
\mathcal{H}(x)\Psi[\sigma]
\end{equation}
where $\mathcal{H}(x)$ is an invariant (scalar) function of the
field quantities \footnote{Schwinger goes on to make a special
choice for $\mathcal{H}(x)$ which yields what he then refers to as
the interaction representation. We do not follow him in this but
stay with general condition (\ref{eq:Schwinger})}. This result
guarantees independence of the evolution with respect to foliation
by a family of space-like hypersurfaces ({$\sigma$). Assuming a
continuous family $\sigma$ ordered by a scalar time-parameter $t$
(only a label!), one may integrate the equation over one of the
surfaces to give
\begin{equation}\label{SE}
i\frac{d\Psi}{dt}=\int_\sigma\sqrt{-g} d^3x\mathcal{H}(x)\Psi
\end{equation}
One may obtain a suitable energy density $\sqrt{-g}\mathcal{H}$,
proceeding classically at first, by considering the action
principle with variation of the boundaries of a four-dimensional
region and the field quantities \cite{Barut},\cite{MTW}. One takes
as action $(I)$
\begin{equation}
I=\int_R\sqrt{-g}d^4xL
\end{equation}
where $L$ is the Lagrangian and $R$ is a four dimensional region
of space-time. Denoting the unit normal 4-vector to the space-like
hypersurface $(\sigma)$ by $n_\mu$, the variation in $R$ may be
written as
\begin{equation}
d\sigma_\mu dx^\nu=\delta\Omega n_\mu n^\nu
\end{equation}
where $dx^\nu$ is a displacement normal to $\sigma$ and
$\delta\Omega$ is the magnitude of the change in four volume.
Hence
\begin{equation}
\delta\int_R\sqrt{-g}d^4xL=\int_R\sqrt{-g}d^4x\delta L+\int_B
Ld\sigma_\mu dx^\nu
\end{equation}
where $B$ is the boundary of $R$. Assuming that the field
equations are satisfied by the field quantities $(\phi^\alpha)$
one gets
\begin{equation}
\delta I=\int_Bd\sigma_\mu\left[\frac{\partial
L}{\partial(\partial_\mu\phi^\alpha)}\delta\phi^\alpha
+\left(g_\nu^\mu L-\partial_\nu\phi^\alpha\frac{\partial
L}{\partial(\partial_\mu\phi^\alpha)}\right)dx^\nu\right]
\end{equation}
which gives
\begin{equation}
\frac{\delta I}{\delta\phi^\alpha}=n_\mu\frac{\partial
L}{\partial(\partial_\mu\phi^\alpha)}=\pi_\alpha
\end{equation}
where $\pi_\alpha$ is the momentum and
\begin{equation}
\frac{\delta
I}{\delta\Omega}=L-\left(n^\nu\partial_\nu\phi^\alpha\right)\pi_\alpha
\end{equation}
and
\begin{equation}
\sqrt{-g}\mathcal{H}=-(\sqrt{-g})\frac{\delta I}{\delta\Omega}
\end{equation}
No restriction on the foliation is required \footnote{One can note
that in terms of the canonical energy-momentum tensor
$T_\tau^\lambda$ one has $\left(-\frac{\delta I}{\delta
\Omega}\right)=n_\lambda T_\tau^\lambda n^\tau$ which agrees with
other authors \cite{Schutz71}, \cite{deWet50} as to the energy
density}.
\section{Real Scalar Field}
In the case of a real scalar field the energy density is given by
\begin{equation}
\sqrt{-g}\mathcal{H}=\sqrt{-g}\frac{1}{2}\left[\left(
n^\lambda\partial_\lambda\phi\right)^2+\partial_j\phi\partial^j\phi+m^2\phi^2\right]
\end{equation}
with $n^\lambda\partial_\lambda\phi=\pi$ the momentum
\footnote{The Latin indices refer to the coordinates on the
space-like hypersurface $\sigma$.}. The metric tensor $g_{\mu\nu}$
adapted to the slicing of space and time in terms of a lapse
function $N$ and shift vectors $N^j$ is given by
\begin{equation}
\left[g_{\mu\nu}\right]=
\left[\begin{array}{ll} N^2-N_SN^S&-N_k \\
-N_i&-^{(3)}g_{ik}
\end{array}
\right]
\end{equation}
and
\begin{equation}
n^\lambda=\left[
\begin{array}{ll} \frac{1}{N}&\frac{-N^k}{N}
\end{array}
\right]
\end{equation}
where $n^\lambda$ is the unit normal to the hypersurface $\sigma$.
When integrating $\mathcal{H}$ over a given hypersurface $\sigma$
the term $\left(\partial_j\phi\partial^j\phi+m^2\phi^2\right)$ may
be replaced by a term $\left(K^\frac{1}{2}\phi\right)^2$ where
\footnote{see appendix}
\begin{equation}
K\phi=-\frac{1}{\sqrt{-g}}\partial_k\left(\sqrt{-g}\partial^k\phi\right)+m^2\phi
\end{equation}
We assume that $K^\frac{1}{2}$ is self-adjoint. Using the usual
wave equation
\begin{equation}
\square\phi+m^2\phi=0
\end{equation}
one gets
\begin{equation}
K\phi=-\frac{1}{\sqrt{-g}}\partial_0\left(\sqrt{-g}\partial^0\phi\right)
\end{equation}
In general one does not have separation of space and time without
restrictions on the metric. A static or stationary metric does
enable such a separation and an analysis in terms of normal modes
\cite{Fulling89}.

\section{Quantization of the Real Scalar Field}
In the representation with $\phi(x)$ a multiplicative operator on
a hypersurface $\sigma$ one introduces the momentum operator $\pi$
which satisfies the covariant commutation relation
\begin{equation}
\left[\phi(x),\pi(y)\right]=i\frac{\delta(x-y)}{\sqrt{-g}}
\end{equation}
(see \cite{Fulling89} for details). One has, therefore,
\begin{equation}
\sqrt{-g}\pi(x)=-i\frac{\delta}{\delta\phi(x)}
\end{equation}
or
\begin{equation}
\pi(x)=-i\frac{1}{\sqrt{-g}}\frac{\delta}{\delta\phi(x)}
\end{equation}
One should especially note that
\begin{equation}
\frac{1}{\sqrt{-g}}\frac{\delta}{\delta\phi(y)}\int\sqrt{-g}d^3xF(x)\phi(x)=F(y)
\end{equation}
For convenience we set
\begin{equation}
K^\frac{1}{2}\phi=\chi
\end{equation}
and
\begin{equation}
-i\frac{1}{\sqrt{-g}}\frac{\delta}{\delta\phi(x)}=-i\frac{\delta}{\delta\phi_g}
\end{equation}
and introduce the operators
\begin{equation}
A^+(x)=\frac{1}{\sqrt{2}}\left(\chi(x)-\frac{\delta}{\delta\phi_g(x)}\right)
\end{equation}
\begin{equation}\label{A}
A^-(x)=\frac{1}{\sqrt{2}}\left(\chi(x)+\frac{\delta}{\delta\phi_g(x)}\right)
\end{equation}
The Hamiltonian is then
\begin{equation}\label{H}
H=\int\sqrt{-g} d^3x A^+(x)A^-(x)
\end{equation}
On flat space-like hypersurfaces this reduces to the conventional
Hamiltonian \cite{Schweber}, \cite{Rider} and also the Hamiltonian
used by Bohm \cite{BHK}. The commutation relation for $\chi(x)$
and $i\pi(y)$ is
\begin{equation}
\left[\chi(x),\frac{\delta}{\delta\phi_g(y)}\right]\Psi=-\frac{\delta\chi(x)}{\delta\phi_g(y)}\Psi
\end{equation}
Since
\begin{equation}
\chi(x)=\left(K^\frac{1}{2}\phi\right)(x)=\int\sqrt{-g}d^3yG(x,y)\phi(y)
\end{equation}
then
\begin{equation}
\frac{\delta \chi(x)}{\delta\phi_g(y)}=G\left(x,y\right)
\end{equation}
One can then write in general
\begin{equation}
\int\sqrt{-g}d^3y\frac{\delta\chi(x)}{\delta\phi_g(y)}h(y)=\left(K^\frac{1}{2}h\right)(x)
\end{equation}
for functions $h(y)$ in the domain of $K^\frac{1}{2}$. A vacuum
state is defined by
\begin{equation}
A^-(x)\Psi_0[\phi]=0 \label{Aminus}
\end{equation}
\begin{equation}
\Psi_0[\phi]=exp\left[-\frac{1}{2}\int\sqrt{-g}d^3x\phi(x)\chi(x)\right]
\end{equation}
One can now form a set of functionals (not normalised)
\begin{eqnarray}\label{Psin}\nonumber
\Psi_n\left[\phi;h_1\ldots h_n\right]=\\
\int\left[\prod_m\left(\sqrt{-g}d^3x_m\right)h_m(x_m)\right]A^+(x_n)A^+(x_{n-1})\ldots
A^+(x_1)\Psi_0[\phi]
\end{eqnarray}
The set of functions $h_1\ldots h_n$ label the functional and have
the same domain as $\phi(x)$. To illustrate this process we
generate $\Psi_1[\phi;h_1]$ and $\Psi_2[\phi;h_1,h_2]$. For
$\Psi_1[\phi;h_1]$ we have
\begin{eqnarray}\nonumber
A^+(z_1)e^{-\frac{1}{2}\int\sqrt{-g}d^3x\phi\chi}\\
\nonumber=\frac{1}{\sqrt{2}}\left(\chi(z_1)-\frac{\delta}{\delta\phi_g(z_1)}\right)e^{-\frac{1}{2}\int\sqrt{-g}d^3x\phi\chi}\\
=\sqrt{2}\chi(z_1)e^{-\frac{1}{2}\int\sqrt{-g}d^3x\phi\chi}
\end{eqnarray}
Hence
\begin{equation}
\Psi_1\left[\phi;h_1\right]=\sqrt{2}\int\sqrt{-g}d^3zh_1(z)\chi(z)e^{-\frac{1}{2}\int\sqrt{-g}d^3x\phi\chi}
\end{equation}
For $\Psi_2[\phi;h_1,h_2]$ we have
\begin{eqnarray}\nonumber
A^+(z_2)A^+(z_1)\Psi_0[\phi]\\
\nonumber=\frac{1}{\sqrt{2}}\left(\chi(z_2)-\frac{\delta}{\delta\phi_g(z_2)}\right)\chi(z_1)e^{-\frac{1}{2}\int\sqrt{-g}d^3x\phi\chi}\\
=\left(\chi(z_2)\chi(z_1)-\frac{\delta\chi(z_1)}{\delta\phi_g(z_2)}+\chi(z_2)\chi(z_1)\right)e^{-\frac{1}{2}\int\sqrt{-g}d^3x\phi\chi}
\end{eqnarray}
Hence
\begin{eqnarray}\nonumber
\Psi_2\left[\phi;h_1,h_2\right]\\
\nonumber=\int\left(\sqrt{-g}d^3z_1\right)\left(\sqrt{-g}d^3z_2\right)A^+(z_2)A^+(z_1)\Psi_0[\phi]\\
\nonumber=2\left(\int\sqrt{-g}d^3xh_2\chi\right)\left(\int\sqrt{-g}d^3xh_1\chi\right)\Psi_0[\phi]\\
-\left(\int\sqrt{-g}d^3xh_1K^{\frac{1}{2}}h_2\right)\Psi_0[\phi]
\end{eqnarray}

As a special case one could consider a hypersurface with a set of
orthogonal functions satisfying
\begin{equation}
K^\frac{1}{2}\phi_n=\omega_n\phi_n
\end{equation}
and then choose the set of $\phi_n$ to specify the functionals.
Putting $\phi=\Sigma q_n\phi_n$ one gets
\begin{eqnarray}
\Psi_1\left[\phi;\phi_n\right]=\sqrt{2}q_n\omega_ne^{-\frac{1}{2}\Sigma
\omega_rq_r^2}\\
\Psi_2\left[\phi;\phi_n,\phi_m\right]
=\left[2q_nq_m\omega_n\omega_m-\omega_m\delta_{nm}\right]e^{-\frac{1}{2}\Sigma
\omega_rq_r^2}
\end{eqnarray}
These are proportional to the usual Hermite functions of the mode
coordinates $q_n$.
\section{Action of the Hamiltonian density on a functional}
Taking the Hamiltonian density from (\ref{H}) and applying it to
the functional of (\ref{Psin}).
\begin{eqnarray}\nonumber
A^+(x)A^-(x)\left[A^+(x_n)A^+(x_{n-1})\ldots
A^+(x_1)\Psi_0\left[\phi\right]\right]\\
\nonumber=A^+(x)\left[A^+(x_n)A^-(x)+\frac{\delta\chi(x)}{\delta\phi_g\left(x_n\right)}\right]A^+\left(x_{n-1}\right)\ldots
A^+\left(x_1\right)\Psi_0\left[\phi\right]
\end{eqnarray}
Repeating the process until $A^-(x)$ operates on
$\Psi_0\left[\phi\right]$ one gets
\begin{eqnarray}\nonumber
\frac{\delta\chi(x)}{\delta\phi_g\left(x_n\right)}A^+(x)A^+(x_{n-1})\ldots
A^+(x_1)\Psi_0[\phi]\\
\nonumber+\frac{\delta\chi(x)}{\delta\phi_g\left(x_{n-1}\right)}A^+(x)A^+(x_n)A^+(x_{n-2})\ldots
A^+(x_1)\Psi_0[\phi]\\
\nonumber+\ldots
\end{eqnarray}
Multiplying this last expression by $h_n\left(x_n\right)\ldots
h_1\left(x_1\right)$ and integrating over $x,x_n,x_{n-1},\ldots
x_1$ one gets
\begin{eqnarray}\nonumber\label{Honf}
\Sigma_m\left(\int\sqrt{-g}d^3x\left(K^\frac{1}{2}h_m\right)(x)A^+(x)\right)\\
\times\Psi_{n-1}\left[\phi;h_nh_{n-1}\ldots h_{m+1},h_{m-1}\ldots
h_1\right]
\end{eqnarray}
We illustrate the above with two examples of $\mathcal{H}$ applied
to $\Psi_1[\phi;h_1]$ and $\Psi_2[\phi;h_1,h_2]$. Firstly, for
$\Psi_1[\phi;h_1]$ we have
\begin{eqnarray}\nonumber
A^+(x)A^-(x)\left[A^+(x_1)\Psi_0\left[\phi\right]\right]\\
\nonumber=A^+(x)\left[A^+(x_1)A^-(x)+\frac{\delta\chi(x)}{\delta\phi_g\left(x_1\right)}\right]\Psi_0\left[\phi\right]\\
\nonumber=\frac{\delta\chi(x)}{\delta\phi_g\left(x_1\right)}A^+(x)\Psi_0\left[\phi\right]\\
\nonumber=\frac{\delta\chi(x)}{\delta\phi_g\left(x_1\right)}\sqrt{2}\chi(x)\Psi_0\left[\phi\right]
\end{eqnarray}
Multiplying by $h_1(x_1)$ and integrating over $x$ and $x_1$ one
gets

\begin{equation}
\sqrt{2}\left[\int\sqrt{-g}d^3x\chi
K^\frac{1}{2}h_1\right]\Psi_0[\phi]
\end{equation}
Secondly, for $\Psi_2[\phi;h_1,h_2]$ we have
\begin{eqnarray}\nonumber
A^+(x)A^-(x)\left[A^+(x_2)A^+(x_1)\Psi_0\left[\phi\right]\right]\\
\nonumber=\left[\frac{\delta\chi(x)}{\delta\phi_g\left(x_1\right)}A^+(x)A^+(x_1)+\frac{\delta\chi(x)}{\delta\phi_g\left(x_1\right)}A^+(x)A^+(x_2)\right]\Psi_0\left[\phi\right]\\
\nonumber=\frac{\delta\chi(x)}{\delta\phi_g\left(x_2\right)}\left[2\chi(x)\chi(x_1)-\frac{\delta\chi(x_1)}{\delta\phi_g\left(x\right)}\right]\Psi_0\left[\phi\right]\\
+\frac{\delta\chi(x)}{\delta\phi_g\left(x_1\right)}\left[2\chi(x)\chi(x_2)-\frac{\delta\chi(x_2)}{\delta\phi_g\left(x\right)}\right]\Psi_0\left[\phi\right]
\end{eqnarray}
multiplying by $h_1(x_1)h_2(x_2)$ and integrating over $x,x_1$ and
$x_2$ one gets
\begin{eqnarray}\nonumber
2\left[\int\sqrt{-g}d^3x\chi
K^\frac{1}{2}h_2\right]\left[\int\sqrt{-g}d^3x\chi
h_1\right]\Psi_0[\phi]\\
\nonumber+2\left[\int\sqrt{-g}d^3x\chi
K^\frac{1}{2}h_1\right]\left[\int\sqrt{-g}d^3x\chi
h_2\right]\Psi_0[\phi]\\
\nonumber
-\left[\int\sqrt{-g}d^3x\left(h_2\right)\left(Kh_1\right)\right]\Psi_0[\phi]\\
-\left[\int\sqrt{-g}d^3x\left(Kh_2\right)\left(h_1\right)\right]\Psi_0[\phi]
\end{eqnarray}

In general one will have to integrate (\ref{SE}) numerically since
the lapse function $(N)$ for the foliation may involve the scalar
time-parameter $t$. Once a solution $\Psi[\sigma]$ is obtained,
the evolution of the Bohm field $\phi_{Bohm}$ on the given
foliation $\sigma$ is determined by the usual Bohmian guidance
condition \cite{BHK}
\begin{equation}\label{BHK}
\frac{1}{N}\frac{d\phi_{Bohm}}{dt}=Im\left[\frac{1}{\Psi^{(t)}[\phi]}\frac{\delta\Psi^{(t)}[\phi]}{\delta\phi_g}\right]
\end{equation}
where successive leaves are labelled by the parameter $t$. In a de
Broglie-Bohm type of theory the foliation should not be subject to
an arbitrary choice but rather be determined by the physical
properties of the system itself. (See section (\ref{beables}) for
further details.)

\section{Integration of the ``Schr\"{o}dinger" equation for static and stationary metric}
In the case of a static or stationary metric one easily obtains
the integrated wave functional in the form:
\begin{equation}
\Psi^{(t=0)}[\phi]=\sum
a_n\Psi_n\left[\phi;h^{(n)}_1\ldots,h^{(n)}_n\right]
\end{equation}
which evolves to
\begin{equation}
\Psi^{(t)}[\phi]=\sum
a_n\Psi_n\left[\phi;e^{-iK^{\frac{1}{2}}t}h^{(n)}_1\ldots,e^{-iK^{\frac{1}{2}}t}h^{(n)}_n\right]
\end{equation}
$\phi$ is the time independent scalar field of the Schr\"{o}dinger
picture. The result follows from the comparison of
$i\frac{d\Psi[\phi]}{dt}$ and equation (\ref{Honf}) which gives
the integral $\int_\sigma\sqrt{-g}d^3x\mathcal{H}(x)\Psi$.

To illustrate the procedure consider a family of space-like
hypersurfaces with a set of orthogonal functions $\phi_n(x)$ such
that
\begin{equation}
K^\frac{1}{2}\phi_n(x)=\omega_n\phi_n(x)
\end{equation}
Let
\begin{equation}
h=\sum b_n\phi_n(x)
\end{equation}
and
\begin{equation}
\phi=\sum c_n\phi_n(x)
\end{equation}
Suppose that at $t=0$
\begin{equation}\label{psi0}
\Psi^{(t=0)}[\phi]=a_0\Psi_0[\phi]+a_1\Psi_1[\phi;h]
\end{equation}
Then
\begin{equation}
\Psi^{(t)}[\phi]=a_0\Psi_0[\phi]+\sqrt{2}a_1\left[\int\sqrt{-g}d^3x\chi
e^{-K^\frac{1}{2}t}h\right]\Psi_0[\phi]
\end{equation}
where
\begin{equation}
\chi=K^\frac{1}{2}\phi=\sum c_n\omega_n\phi_n(x)
\end{equation}
Therefore
\begin{equation}
\Psi^{(t)}[\phi]=a_0\Psi_0[\phi]+\sqrt{2}a_1\left[\sum
b_nc_n\omega_ne^{-i\omega_n t}\right]\Psi_0[\phi]
\end{equation}
$\phi_{Bohm}$ will be time dependent so we write
\begin{equation}
\phi_{Bohm}=\sum q_n(t)\phi_n
\end{equation}
Noting that from the definition of the vacuum, (\ref{A}),
(\ref{Aminus})
\begin{equation}
\frac{\delta\Psi_0[\phi]}{\delta\phi_g(x)}=-\chi(x)\Psi_0[\phi]
\end{equation}
we have
\begin{equation}
\left[\frac{1}{\Psi^{(t)}[\phi]}\frac{\delta\Psi^{(t)}[\phi]}{\delta\phi_g}\right]=-\chi(x)+\frac{\sqrt{2}a_1\left[\sum
b_n\omega_n
e^{-i\omega_nt}\phi_n(x)\right]}{a_0+\sqrt{2}a_1\left[\sum
b_nc_n\omega_ne^{-i\omega_nt} \right]}
\end{equation}
Since in this example $\chi$ is a real field, the guidance
condition, equation (\cite{BHK}), yields
\begin{equation}\label{phib}
\frac{1}{N}\frac{d\phi_{Bohm}}{dt}=Im\left[\frac{1}{\Psi^{(t)}[\phi]}\frac{\delta\Psi^{(t)}[\phi]}{\delta\phi_g}\right]=Im\left[\frac{\sqrt{2}a_1\left[\sum
b_n\omega_n
e^{-i\omega_nt}\phi_n(x)\right]}{a_0+\sqrt{2}a_1\left[\sum
b_nc_n\omega_ne^{-i\omega_nt} \right]}\right]
\end{equation}
The $c_n$ in equation (\ref{phib}) must, of course, be the
coefficients giving $\phi_{Bohm}$ at time $t$.
\section{Beables of the Field}\label{beables}
In constructing a theory of ``beables"\footnote{The concept of
beables was introduced by J. S. Bell \cite{Bell1987}.}  one is
free to choose the initial space-like hypersurface and field
$\phi_{Bohm}$ thereon. Each choice corresponds to a different
state of affairs. However, in integrating equation (\ref{BHK}) one
needs to give an invariant rule constructing the next leaf of the
foliation in order to get a definite $\phi_{Bohm}$. One
restriction is to choose to displace each point on the initial
surface by equal proper times; the direction of the displacement
will also need to be specified by a unit time-like four-vector
$[W_\mu]$. The construction of the foliation is shown in figure 1.
The shift vectors are chosen to be zero as other choices only
correspond to different coordinate systems on the surface
\begin{equation}
ds^2=N^2dt^2-dx^ig_{ij}dx^j
\end{equation}
The three velocity of the shift given by $[W^\mu]$ is
\begin{equation}
v^i=\frac{W^i}{W^0}
\end{equation}
the proper time along the shift vector is $d\epsilon$, constant
for all points on the surface and the proper time along the normal
is $Ndt$. Therefore
\begin{equation}
Ndt=d\epsilon cosh\theta
\end{equation}
with
\begin{equation}
cosh\theta=\frac{1}{\sqrt{1-v^2}}
\end{equation}
It is convenient to choose $N=cosh\theta$ so that $dt=d\epsilon$.
In one spatial dimension one also has
\begin{equation}
g_{11}=sinh\theta=\frac{v}{\sqrt{1-v^2}}
\end{equation}
or, in general
\begin{equation}
dx^ig_{ij}dx^j=\left(\frac{v}{\sqrt{1-v^2}}d\epsilon\right)^2
\end{equation}
for displacements along the vector $[W^\mu]$. If
\begin{equation}
\frac{1}{N}\frac{d\phi_{Bohm}}{dt}=\frac{\delta S}{\delta \phi_g}
\end{equation}
on one leaf of the foliation at $t$ then on the leaf $t+dt$
\emph{at the same coordinate point} $[x^i]$
\begin{equation}
\Delta\phi=\frac{d\phi}{dt}\cdot d\epsilon=N\frac{\delta S}{\delta
\phi_g}d\epsilon
\end{equation}
One has interpreted $\frac{1}{N}\frac{d\phi}{dt}$ as the momentum
and, writing $\Psi[\phi]=R[\phi]e^{iS[\phi]}$, replaced it by
$\frac{\delta S}{\delta \phi_g}$ as in Bohm's original theory.

One possible choice of $W^\mu$ is to follow the rule we have
previously adopted in the many-particle case and use the time-like
eigenvector of the energy-momentum tensor $T_{\mu\nu}$
\cite{Horton}, \cite{horton2002}, \cite{dewdney}. The $T_{\mu\nu}$
will be that of a classical scalar field $\phi$ but with $\phi$
replaced by $\phi_{Bohm}$.

\section{Appendix}
\begin{equation}
\int\sqrt{-g}d^3x\left[\frac{1}{\sqrt{-g}}\frac{\partial}{\partial
x^k}\left(\sqrt{-g}\phi\partial^k\phi\right)\right] =\int
d^3x\frac{\partial}{\partial
x^k}\left(\sqrt{-g}\phi\partial^k\phi\right)
\end{equation}
This expression can be converted into an integral over a bounding
surface which will vanish for suitable behaviour of
$\sqrt{-g}\phi\partial^k\phi$. Hence
\begin{equation}
\int\sqrt{-g}d^3x\left[\partial_k\phi\partial ^k\phi+\phi
\left(\frac{1}{\sqrt{-g}}\partial_k\left(\sqrt{-g}\partial^k\phi\right)\right)\right]=0
\end{equation}
and
\begin{equation}
\int\sqrt{-g}d^3x\left[\partial_k\phi\partial
^k\phi\right]=-\int\sqrt{-g}d^3x\left[\phi
\left(\frac{1}{\sqrt{-g}}\partial_k\left(\sqrt{-g}\partial^k\phi\right)\right)\right]
\end{equation}

\section{Figure Captions}

\begin{description}
\item  Figure 1. The construction of the foliation.

\end{description}

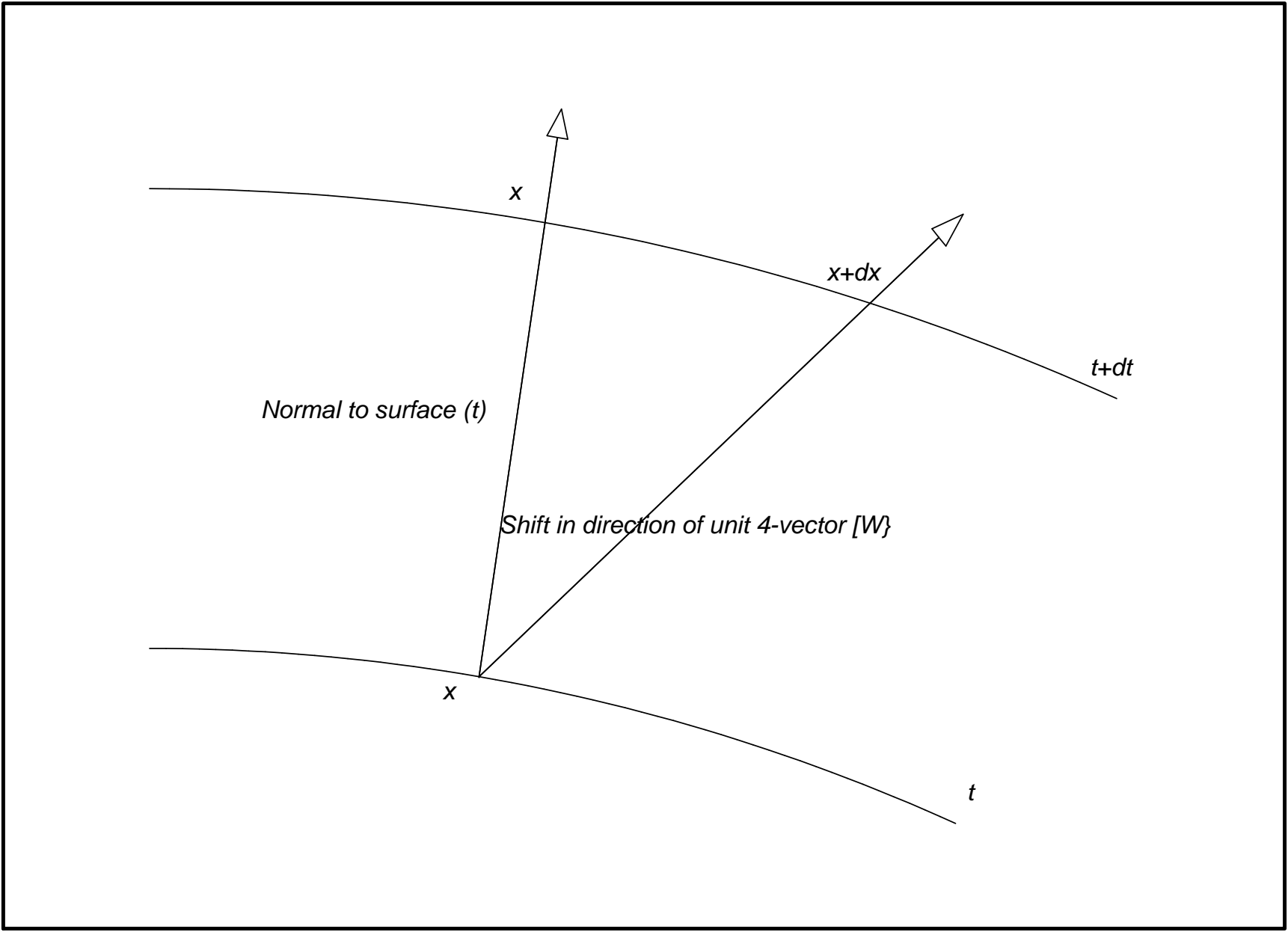
\end{document}